\documentclass[11pt]{article}
\usepackage[top=0.9in, bottom=0.9in, left=0.8in, right=0.8in]{geometry}
\usepackage{setspace}
\usepackage[T1]{fontenc}
\usepackage{times}
\usepackage{booktabs}
\usepackage{rotating}
\usepackage{bbm}
\usepackage{dsfont}
\usepackage{graphicx}
\usepackage{multicol}
\usepackage[section]{placeins} 
\usepackage[large, bf]{caption}
\usepackage[FIGTOPCAP]{subfigure}
\usepackage{pdfpages}
\usepackage{palatino}
\usepackage{pdflscape}
\usepackage{textcomp}
\usepackage{nicefrac}
\usepackage{geometry}
\usepackage{float}
\usepackage{changepage}
\usepackage{makecell}
\usepackage{array}
\usepackage{booktabs} 
\usepackage{threeparttable}
\usepackage{tabularx}

\usepackage{natbib}
\bibpunct{(}{)}{;}{a}{,}{,}

\usepackage{eurosym,ulem,color,sectsty,comment,footmisc,subfiles}

\usepackage[toc,page]{appendix}
\usepackage{afterpage}
\usepackage{changepage}
\usepackage{booktabs}

\newcolumntype{L}[1]{>{\raggedright\let\newline\\arraybackslash\hspace{0pt}}m{#1}}
\newcolumntype{C}[1]{>{\centering\let\newline\\arraybackslash\hspace{0pt}}m{#1}}
\newcolumntype{R}[1]{>{\raggedleft\let\newline\\arraybackslash\hspace{0pt}}m{#1}}

\usepackage{amsmath, amsfonts, amssymb, amsthm}

\usepackage{mathpazo} 
\usepackage[pdftex]{hyperref}
\hypersetup{colorlinks, citecolor=black, filecolor=blue, linkcolor=blue, urlcolor=blue}

\def\urltilda{\kern -.15em\lower .7ex\hbox{\~{}}\kern .04em}

\usepackage{titlesec}


\titlespacing*{\section}{0pt}{1.5ex plus 1ex minus .2ex}{0.8ex plus .2ex}
\titlespacing*{\subsection}{0pt}{1.2ex plus 1ex minus .2ex}{0.8ex plus .2ex}
\titlespacing*{\subsubsection}{0pt}{1ex plus 1ex minus .2ex}{0.8ex plus .2ex}

\begin{document}

\title{Online Appendix \& Additional Results for ``The Determinants of Social Connectedness in Europe''}
\author{\hspace{0.8cm} Michael Bailey\thanks{Facebook. Email: \href{mailto:mcbailey@fb.com}{mcbailey@fb.com}} \and
	\hspace{0.8cm}
	Drew Johnston\thanks{Harvard University. Email: \href{mailto:drewjohnston@g.harvard.edu}{drewjohnston@g.harvard.edu}}\and
	\hspace{0.8cm}
	Theresa Kuchler\thanks{New York University, Stern School of Business; CEPR. Email: \href{mailto:tkuchler@stern.nyu.edu}{tkuchler@stern.nyu.edu}} \and
	\hspace{0.8cm}
	Dominic Russel\thanks{New York University, Stern School of Business. Email: \href{mailto:drussel@stern.nyu.edu}{drussel@stern.nyu.edu}} \and 
	\hspace{0.8cm}
	Bogdan State\thanks{Facebook. Email: \href{mailto:bogdanstate@fb.com}{bogdanstate@fb.com}} \and
	\hspace{0.8cm}
	Johannes Stroebel\thanks{New York University, Stern School of Business; NBER; CEPR. Email: \href{mailto:jcstroebel@gmail.com}{johannes.stroebel@nyu.edu}}\vspace{0.2cm}
		}
\date{July 23, 2020}
\maketitle

\normalsize

\section{Additional Case Studies of European Social Connectedness} \label{appendix:case_studies}

\begin{figure}[hp]
  \caption{Social Network Distributions in Belgium}
  \label{fig:belgium}
  \makebox[\textwidth][c]{\includegraphics[width=0.9\textwidth]{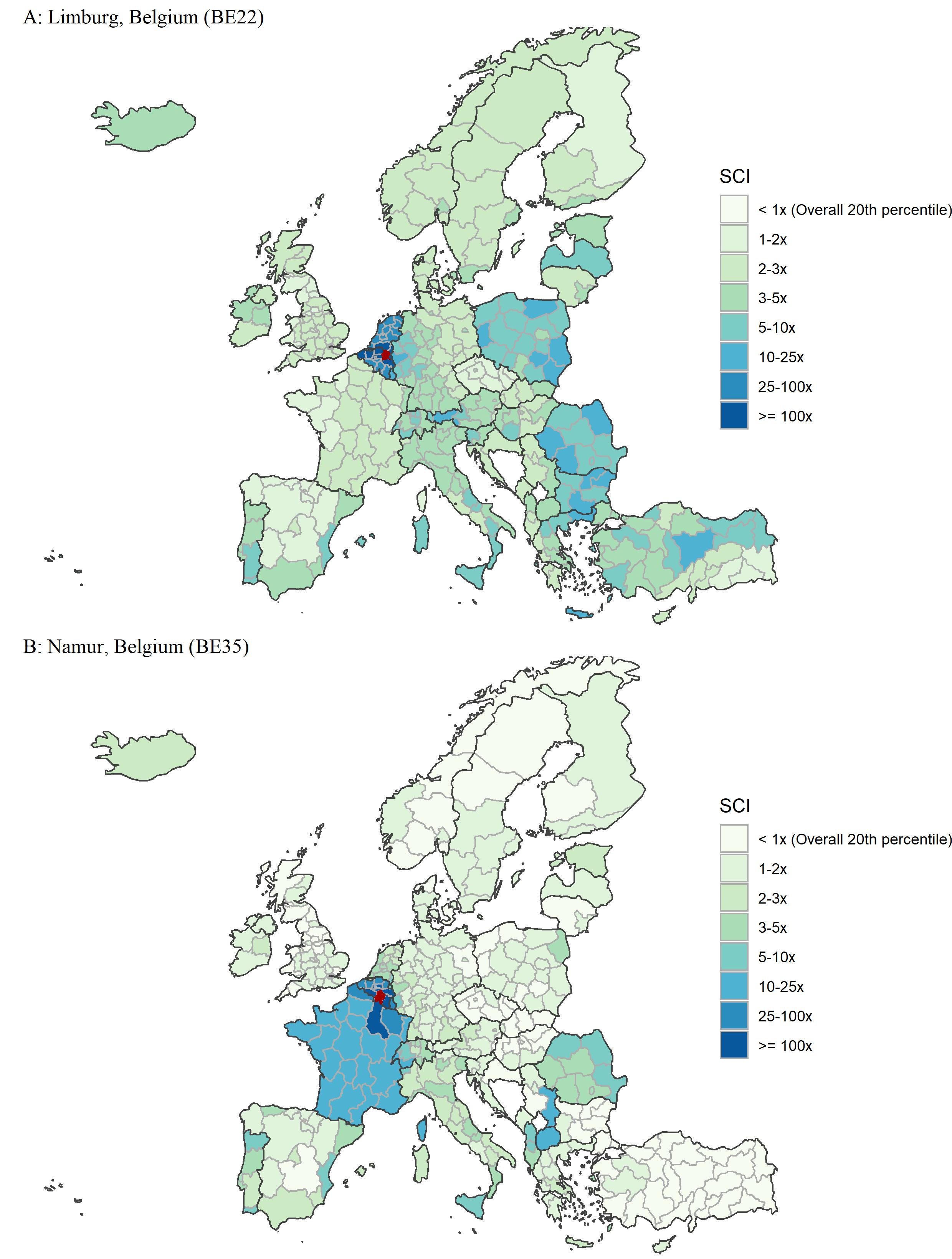}}
    \begin{footnotesize}
        \textit{Note:} Figure shows the relative probability of connection, measured by $SocialConnectedness_{ij}$, of all European regions $j$ with two regions $i$: Limburg, BE (Panel A) and Namur, BE (Panel B). The measures are scaled from the 20th percentile of all $i,j$ pairs in Europe. Darker regions have a higher probability of connection.
    \end{footnotesize}
\end{figure}

\paragraph*{Belgium.} Figure \ref{fig:belgium} maps the social network of two regions in Belgium: Limburg, in Panel A, and Namur, in Panel B. Again, both regions are most strongly connected to nearby regions within their own country. Yet, while the capitals of the two regions (Hasslet and Namur, respectively) are less than 70km apart, the two regions' connections outside Belgium differ substantially. The official and most commonly spoken language in Limburg is Dutch, whereas in Namur it is French. Accordingly, Limburg is more strongly connected to the entire Netherlands to the north and Namur is more strongly connected to areas throughout all of France to the south.

\begin{figure}[hp]
  \caption{Social Network Distributions in Switzerland}
  \label{fig:switzerland}
  \makebox[\textwidth][c]{\includegraphics[width=0.9\textwidth]{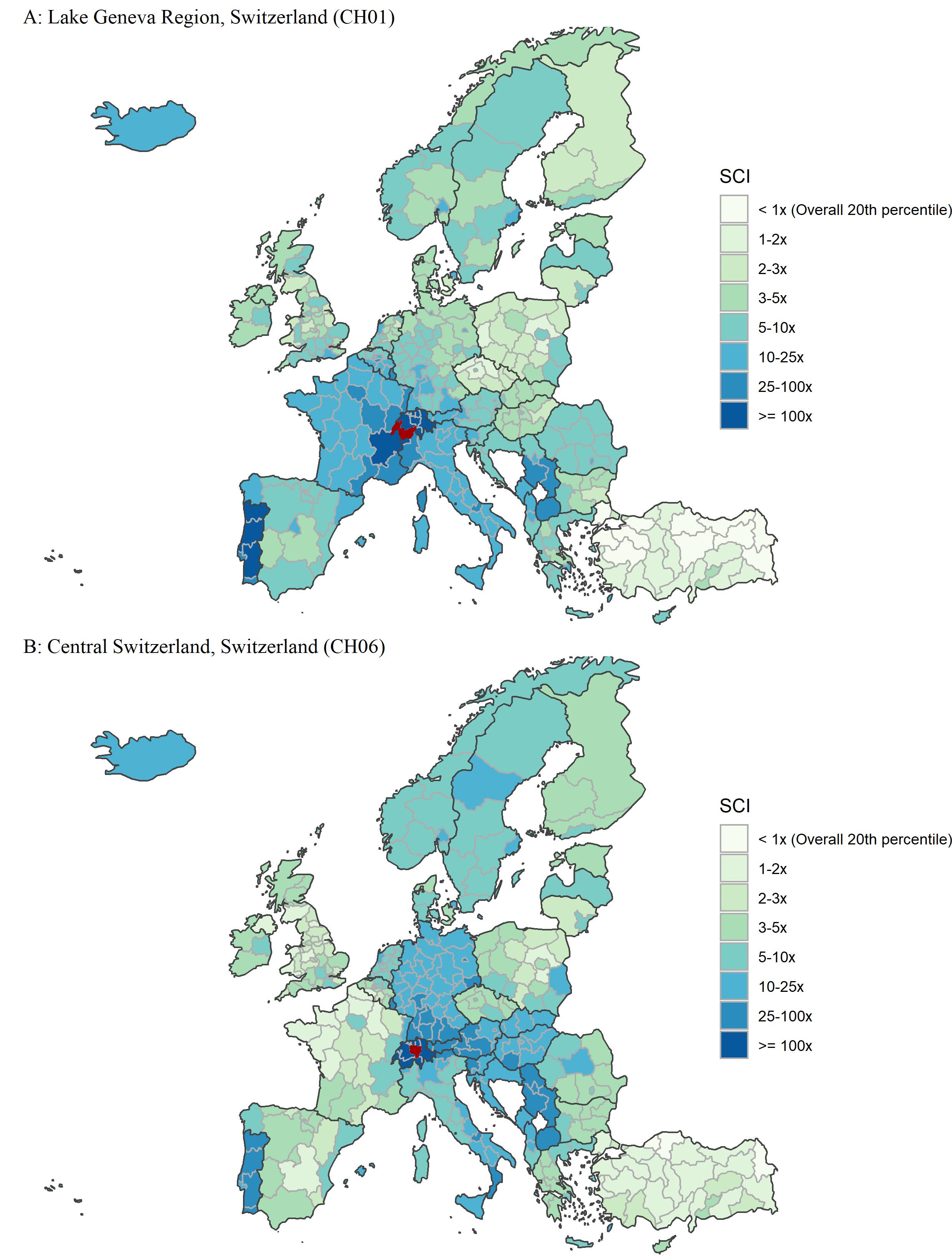}}
    \begin{footnotesize}
        \textit{Note:} Figure shows the relative probability of connection, measured by $SocialConnectedness_{ij}$, of all European regions $j$ with two regions $i$: Lake Geneva Region, CH (Panel A) and Central Switzerland, CH (Panel B). The measures are scaled from the 20th percentile of all $i,j$ pairs in Europe. Darker regions have a higher probability of connection.
    \end{footnotesize}
\end{figure}

\paragraph*{Lake Geneva and Central Switzerland.} Figure \ref{fig:switzerland} shows the social connectedness of two neighboring regions in Switzerland, the Lake Geneva Region (Panel A) and Central Switzerland (Panel B). The Lake Geneva Region includes the Swiss cantons of Geneva, Vaud, and Valais, all of which are primarily French speaking. Central Switzerland and includes the cantons of Lucerne, Uri, Schwyz, Obwalden, Nidwalden, and Zug, all of which are primarily German speaking. Accordingly, the connections of the Lake Geneva Region are strong throughout all of France to the west, whereas the connections of Central Switzerland are stronger throughout all of Germany to the north. These patterns provide more evidence of language playing an important role in shaping social connectedness.

\begin{figure}[hp]
  \caption{Social Network Distributions in Turkey}
  \label{fig:turkey}
  \makebox[\textwidth][c]{\includegraphics[width=0.9\textwidth]{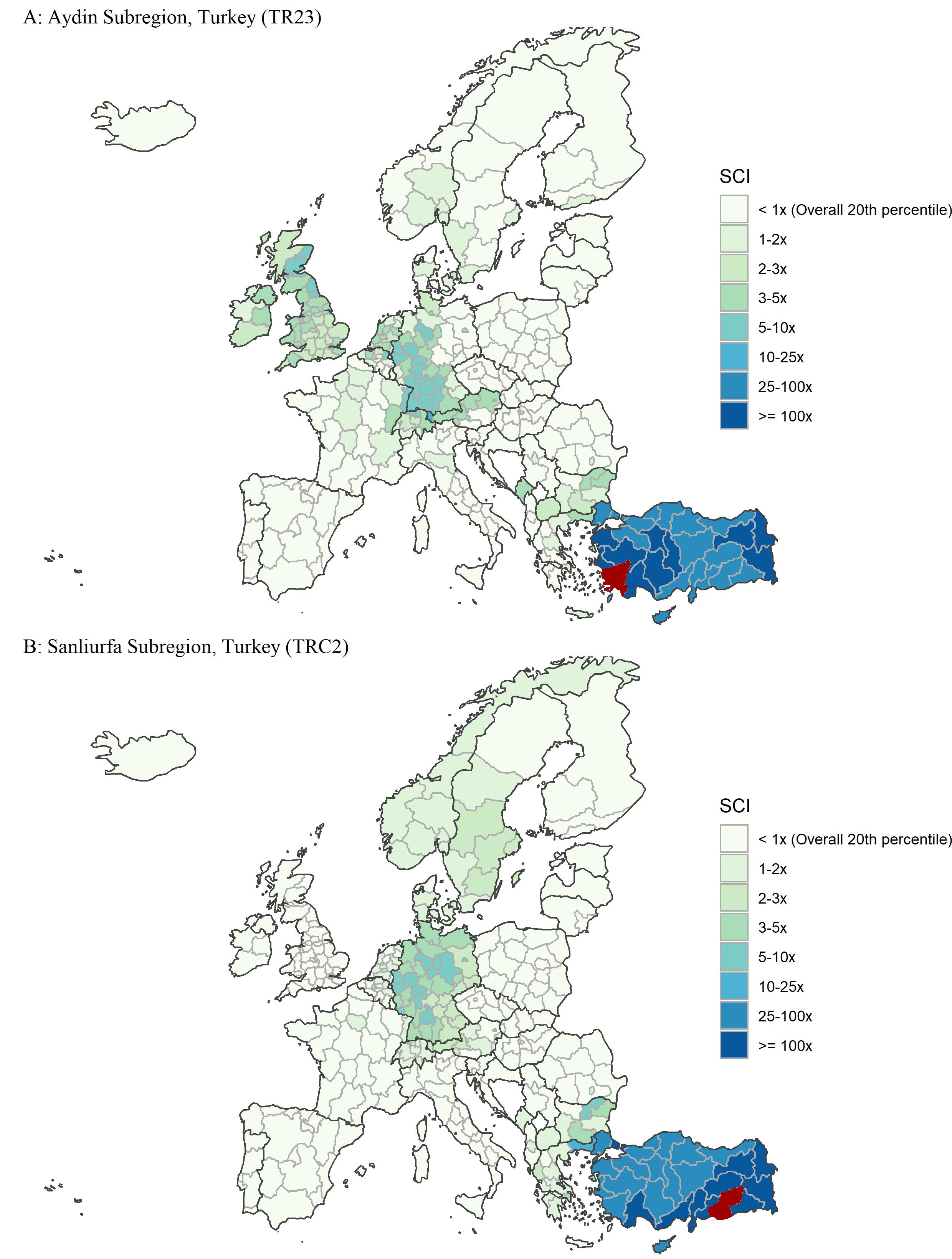}}
    \begin{footnotesize}
        \textit{Note:} Figure shows the relative probability of connection, measured by $SocialConnectedness_{ij}$, of all European regions $j$ with two regions $i$: the Ayd{\i}n Subregion, TR (Panel A) and the \c{S}anl{\i}urfa Subregion, TR (Panel B). The measures are scaled from the 20th percentile of all $i,j$ pairs in Europe. Darker regions have a higher probability of connection.
    \end{footnotesize}
\end{figure}

\paragraph*{Ayd{\i}n and \c{S}anl{\i}urfa.} Figure \ref{fig:turkey} shows the connections of two regions in Turkey, the Ayd{\i}n Subregion (Panel A) and the \c{S}anl{\i}urfa Subregion (Panel B). Ayd{\i}n's connections to Germany appear similar to the Samsun Subregion (shown in the main paper), with connectedness dropping-off at the former East German border. This pattern can be seen in most Turkish regions, likely reflecting West Germany's 1961-1973 labor recruitment agreement with Turkey. However, there is no similar drop-off for the \c{S}anl{\i}urfa Subregion in Panel B, which shows relatively strong connections to all of Germany. The \c{S}anl{\i}urfa Subregion is home to a sizeable Kurdish population, including the largely Kurdish city of Diyarbakr. Armed conflict between the Kurdish Workers' Party and the Turkish government in the 1990s led many Kurds to seek refuge in Germany \citep{turkishMigration}. Much of this migration happened after the reunification of East and West Germany, likely explaining the \c{S}anl{\i}urfa's connectedness to the entire country.

\begin{figure}[hp]
  \caption{Social Network Distributions in France}
  \label{fig:france}
  \makebox[\textwidth][c]{\includegraphics[width=0.9\textwidth]{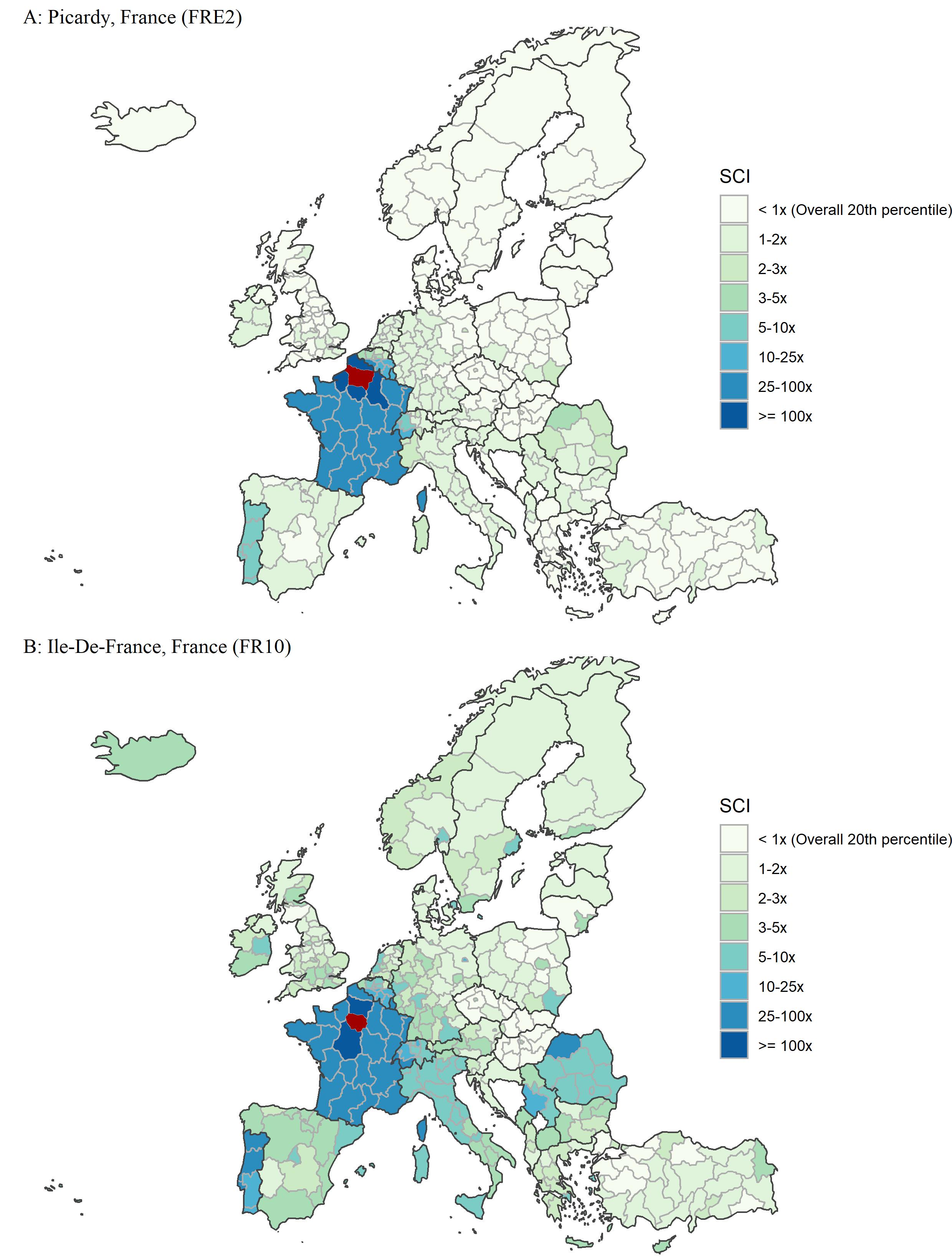}}
    \begin{footnotesize}
        \textit{Note:} Figure shows the relative probability of connection, measured by $SocialConnectedness_{ij}$, of all European regions $j$ with two regions $i$: Picardy, FR (Panel A) and \^{I}le-de-France (Panel B). The measures are scaled from the 20th percentile of all $i,j$ pairs in Europe. Darker regions have a higher probability of connection.
    \end{footnotesize}
\end{figure}

\paragraph*{Picardy and \^{I}le-de-France.} Figure \ref{fig:france} shows the connections of two neighboring regions in France, Picardy (Panel A) and \^{I}le-de-France (Panel B). Picardy is most strongly connected to other regions in France, with connectedness dropping off sharply at country borders. Indeed, only 6.2\% of the region's European connections are to individuals located outside of France. By contrast, \^{I}le-de-France, which includes Paris and its metropolitan area, shows fairly strong connections to a number of regions outside of France---17.3\% of the region's European connections are international. In particular, \^{I}le-de-France's connections to regions with other capital cities, including Madrid, Berlin, Prague, Warsaw, Oslo, and Stockholm, appear comparatively much stronger. This is an example of the differences in international connectedness between European regions.

\begin{figure}[hp]
  \caption{Social Network Distributions in Continental Croatia (HR04)}
  \label{fig:croatia}
  \makebox[\textwidth][c]{\includegraphics[width=0.9\textwidth]{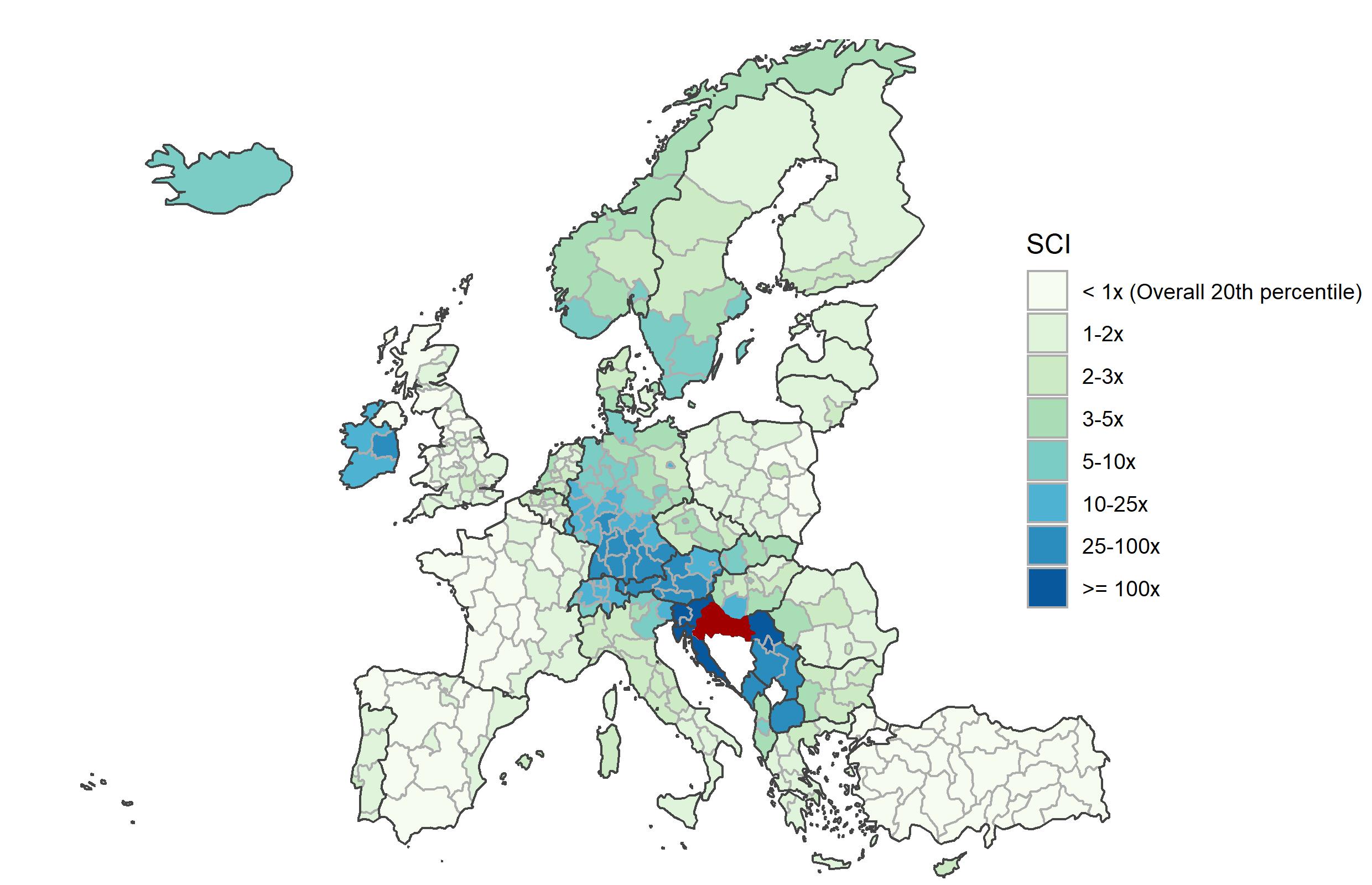}}
    \begin{footnotesize}
        \textit{Note:} Figure shows the relative probability of connection, measured by $SocialConnectedness_{ij}$, of all European regions with Continental Croatia. The measures are scaled from the 20th percentile of all $i,j$ pairs in Europe. Darker regions have a higher probability of connection.
    \end{footnotesize}
\end{figure}

\paragraph*{Continental Croatia.} Figure \ref{fig:croatia} shows the connectedness of Continental Croatia, one of Croatia's two NUTS2 regions. The connections of the region are strongest with other countries that were part of Yugoslavia---including Slovenia, Serbia, Montenegro, and North Macedonia---suggestive of the importance of past country borders in shaping connectedness. There are also strong connections between Croatia and Ireland, likely related to a pattern of migration since Croatia's 2013 accession into the European Union \citep[see, for example,][]{immigration_croatia2019, immigration_croatia2017}. Similarly, there are connections to central European countries, which are geographically close and have also welcomed Croatian migrants.

\section{Agglomerative Clustering Algorithm}

\begin{figure}[hp]
    \caption{Socially Connected Communities within Europe}
    \label{fig:clusters}
    \makebox[\textwidth][c]{\includegraphics[width=0.9\textwidth]{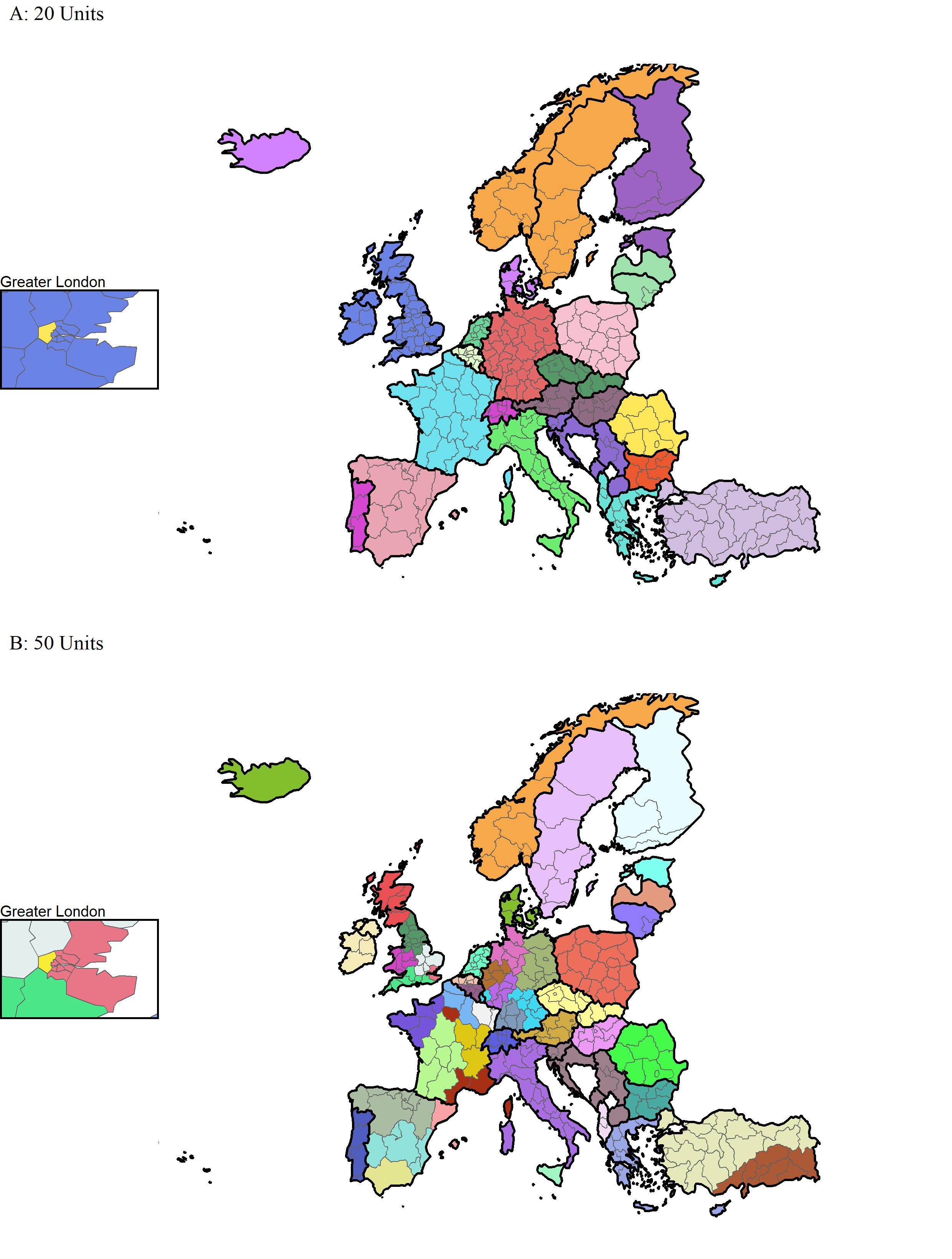}}
        \begin{footnotesize}
        \textit{Note:} Figure shows European regions grouped together when we use hierarchical agglomerative linkage clustering to create 20 (Panel A) and 50 (Panel B) distinct groups of regions.
    \end{footnotesize}
\end{figure}

The previous examples suggest that social connections are substantially stronger within countries than between countries, but that there is also substantial within-country variation based on certain regional characteristics. We next seek to understand how these patterns would be reflected if we created communities of regions with strong connections to each other. To do so, we create clusters that maximize within-cluster pairwise social connectedness using hierarchical agglomerative linkage clustering.\footnote{There are a number of possible algorithms that could construct such clusters. Conceptually, the agglomerative clustering algorithm starts by considering each of the N countries as a separate community of size one. It then iteratively combines clusters that are ``closest'' together. Here, our measure of distance is $1/SocialConnectedness_{i,j}$. The distance between clusters is the average of the pairwise distances between regions in the clusters.} Figure \ref{fig:clusters} shows the results when we use this algorithm to group Europe into 20 and 50 communities, instead of the 37 existing countries.

In Panel A, the 20 unit map, nearly all of the community borders (denoted by a change in area color) line up with country borders (denoted by large black lines), consistent with strong intra-country social connectedness. The only exception is in the United Kingdom, where one region, Outer London West and North West, is grouped together with Romania. This area of London has welcomed a large number of Romanian immigrants in recent years and includes Burnt Oak, a community nicknamed ``Little Romania'' \citep{romaniaLondon}. Furthermore, the borders of cross-country communities mostly line up with historical political borders. For example, every region in the countries that made up Yugoslavia until the early 1990s for which we have data---Slovenia, Croatia, Serbia, North Macedonia, and Montenegro---are grouped together in one community. The same is true of the regions in the Czech Republic and Slovakia, which were united as Czechoslovakia until 1993. Other cross-country communities line up with even older historical borders: the United Kingdom and Ireland, Denmark and Iceland, Sweden and Norway, and Austria and Hungary all split from political unions in the first half of the 20th century but remain grouped together by present-day social connectedness.

While much of the alignment between country and community borders persists in the 50 unit map in Panel B of Figure \ref{fig:clusters}, countries begin to break apart internally. Most of these resulting sub-country communities are spatially contiguous, consistent with distance playing a strong role in social connectedness within countries. One notable exception is \^{I}le-de-France, which is home to Paris and accounts for nearly 30\% of French GDP. The region is grouped with France's southeastern coast (often referred to as French Riviera) and Corsica, popular vacation destinations for the well-heeled. We also see linguistic communities form: Belgium splits into a French and a Dutch speaking community, and Catalan and Andalusian Spanish communities emerge in Spain. The southeastern corner of Turkey, which has a much higher concentration of Kurds than the rest of the country, also forms its own community. Finally, historical country borders remain influential: the regions of former Czechoslovakia, former Yugoslavia, and Denmark and Iceland still form single communities, while former East Germany forms its own community.

\section{Social Connectedness and Travel} \label{sec:travel}

Social connectedness between two regions may be related to certain economic and social interactions. For example, \cite{Bailey2018b} documents correlations between social connectedness, trade flows, patent citations, and migration patterns between U.S. counties. \cite{Bailey2019a} looks more specifically at transportation, highlighting the relationship between transportation infrastructure and urban social networks. In this section, we look at the relationship between social connectedness and European passenger train flows. Our unique data allow us to add a continental-level empirical analysis to an existing literature of theoretical and survey-based studies that explore the influence of social networks on travel decisions \citep[for an overview, see][]{kim2018networks}.

\paragraph*{Regional Data on European Travel.} For our dependent variable, we use the number of passenger train trips between each regional pair $i$ and $j$. We discuss the availability of these data, as well as our procedure for cleaning and standardization, in Appendix \ref{appendix:pass_train}. We also use information on rail and drive travel times between region geographic centers. Rail travel times come from the European Transport Information System 2010 ``observed'' data.\footnote{Data available at: \url{http://ftp.demis.nl/outgoing/etisplus/datadeliverables/TextFiles/}.} Drive times were generated using the Open Source Routing Machine, an OpenStreetMap-based routing service.\footnote{OSRM relies on only open-source software and does not use real-time data (e.g., traffic). Both of these factors increase replicability compared to other online-mapping services. For more information, see: \url{http://project-osrm.org/}.}

\paragraph*{Relationship between Travel and Connectedness.} As the outcome variable is non-negative and contains many zeros, we follow \cite{Correia} and estimate a Poisson Pseudo-Maximum Likelihood regression model. Specifically, our equation of interest is:
\begin{equation}
\begin{split}
    \label{eq:reg_train}
    PassengerTrain_{ij} = exp[\beta_0\log(SocialConnectedness_{ij}) + X_{ij} + \alpha\log(D_{ij}) + \psi_{i} + \psi_{j}] \cdot \epsilon_{ij}
\end{split}
\end{equation}
Here, the definitions of log($SocialConnectedness_{ij}$), $X_{ij}$, $\psi_{i}$, and $\psi_{j}$ remain unchanged from Equation 2 in the primary paper. The variable $PassengerTrain_{ij}$ is the number of rail passengers that travel from region $i$ to region $j$ in a given year. The vector for ``distance,'' represented by $D_{ij}$, includes the geographic distance, as well as the rail and driving travel times in minutes between the central points of the regions.

Table \ref{tab:passenger_train} shows the results from Regression \ref{eq:reg_train}. Due to differences in data availability between years, we present results using the most recent year for which passenger train flows for a given $i,j$ pair is available (in columns 1-5) and, separately, results using 2015, 2010, and 2005 (columns 6, 7, and 8). Column 1 includes the log of $SocialConnectedness_{ij}$ and the NUTS2 region fixed effects as the only explanatory variables. We find that a 10\% increase in connectedness between a pair of regions is associated with a 17\% increase in passenger rail traffic between the two. Column 2 adds the log of the geographic distance between the regions, which, intuitively, has a negative relationship with the number of passenger train trips. In column 3, we add controls for the travel time, separately by train and car, between the central points of the regions. In general, we find that these travel times have a stronger negative relationship with passenger rail traffic than distance alone. Column 4 adds all country pair fixed effects; column 5 adds all the demographic/soceioeconomic controls from Table 1 in the primary paper; and columns 6, 7, and 8 repeat these analyses for the years 2015, 2010, and 2005, respectively. Adjusting for geographic distance, travel time by car and train, and country fixed effects, we find that, on average, a pair of regions with 10\% higher social connectedness will have between 11.7 and 13.6\% more rail passengers travel between them. Overall, our results provide large-scale empirical evidence to support existing models that suggest social networks play an important role in individuals' travel decisions \citep[see e.g.,][]{axhausen2008networks, carrasco2009social, paez2007social}.

\begin{table}
\begin{threeparttable}[h]
    \begin{singlespacing}
    \begin{footnotesize}
    \caption{Social Connectedness and Passenger Train Travel}
    \label{tab:passenger_train}
    \footnotesize
    {
    \def\sym#1{\ifmmode^{#1}\else\(^{#1}\)\fi}
    \begin{tabularx}{\textwidth}{l*{8}{>{\centering\arraybackslash} X }}
        \toprule\toprule
            &\multicolumn{8}{c}{Dependent Variable: log(Passenger Rail Trips)}  \\       \midrule
            &\multicolumn{1}{c}{(1)}         &\multicolumn{1}{c}{(2)}         &\multicolumn{1}{c}{(3)}         &\multicolumn{1}{c}{(4)}         &\multicolumn{1}{c}{(5)}         &\multicolumn{1}{c}{(6)}         &\multicolumn{1}{c}{(7)}         &\multicolumn{1}{c}{(8)}         \\
            & Most recent         & Most recent         & Most recent         & Most recent         & Most recent         &        2015         &        2010         &        2005         \\
        \midrule
        log(SocialConnectedness)&       1.711\sym{***}&       1.506\sym{***}&       1.416\sym{***}&       1.290\sym{***}&       1.364\sym{***}&       1.241\sym{***}&       1.167\sym{***}&       1.236\sym{***}\\
                    &     (0.052)         &     (0.080)         &     (0.101)         &     (0.112)         &     (0.099)         &     (0.056)         &     (0.071)         &     (0.093)         \\
                    &                     &                     &                     &                     &                     &                     &                     &                     \\
        log(Distance in KM)&                     &      -0.361\sym{**} &       0.259         &       0.041         &       0.118         &       0.109         &       0.627\sym{***}&       0.076         \\
                    &                     &     (0.180)         &     (0.327)         &     (0.354)         &     (0.284)         &     (0.172)         &     (0.198)         &     (0.252)         \\
                    &                     &                     &                     &                     &                     &                     &                     &                     \\
        log(Rail Time in Mins)&                     &                     &      -0.493         &      -0.004         &       0.104         &      -1.096\sym{***}&      -1.071\sym{***}&       0.036         \\
                    &                     &                     &     (0.442)         &     (0.528)         &     (0.543)         &     (0.162)         &     (0.251)         &     (0.490)         \\
                    &                     &                     &                     &                     &                     &                     &                     &                     \\
        log(Drive Time in Mins)&                     &                     &      -0.396\sym{**} &      -0.446\sym{**} &      -0.448\sym{***}&      -0.013         &      -0.562\sym{***}&      -0.409\sym{**} \\
                    &                     &                     &     (0.201)         &     (0.198)         &     (0.170)         &     (0.110)         &     (0.172)         &     (0.165)         \\
                    &                     &                     &                     &                     &                     &                     &                     &                     \\
        NUTS2 FEs   &           Y         &           Y         &           Y         &           Y         &           Y         &           Y         &           Y         &           Y         \\
        All Country Pair FEs&                    &                    &                    &           Y         &           Y         &           Y         &           Y         &           Y         \\
        Table 1 Controls&                    &                    &                    &                    &           Y         &           Y         &           Y         &           Y         \\
        \midrule
        pseudo-$R^2$&       0.964         &       0.965         &       0.965         &       0.972         &       0.975         &       0.980         &       0.980         &       0.975         \\
        Number of Observations&      85,390         &      85,390         &      77,390        &      77,390         &      61,810         &       27,650         &      32,176         &      58,462         \\
        N Explained by FEs&      30,698         &      30,698         &      23,784        &      49,000         &      36,517         &       19,146         &      16,269         &      29,659         \\
        \bottomrule\bottomrule
    \end{tabularx}
    }
    \begin{tablenotes}[flushleft]
    \item[] \textit{Note:} Table shows results from Regression \ref{eq:reg_train}. The unit of observation is a NUTS2 region pair. The dependent variable in all columns is the number of passenger rail trips in 2015, 2010, or 2005 from region $i$ to region $j$. In columns 1-5, we use the most recent year for which these data are available for a given $i,j$ pair. In columns 6, 7, and 8, we use only the 2015, 2010, and 2005 data, respectively. Column 1 shows the results from using the log of $SocialConnectedness_{ij}$ and NUTS2 region fixed effects as the only explanatory variables. Column 2 adds the log of the geographic distance between the regions. Column 3 adds the log of the travel time, by train and car, between the central points of the regions. Column 4 adds all country-pair fixed effects. Columns 5-8 add the demographic/soceioeconomic controls from Table 1 in the primary paper. Observations that are fully explained by fixed effects are dropped before the PPML estimation. Standard errors are double clustered by each region $i$ and region $j$ in a region pair. Significance levels: *(p$<$0.10), **(p$<$0.05), ***(p$<$0.01).
    \end{tablenotes}
    \end{footnotesize}
    \end{singlespacing}
\end{threeparttable}
\end{table}

\section{International Connections and Views on European Union} \label{sec:eu_views}

A central goal of the European Union is to enhance cohesion and solidarity across European countries, and a variety of programs, such as the Erasmus exchange student program, explicitly exist to foster this connectivity \citep{eu_goals, erasmus}. However, in recent years there has been a decline in trust in the EU and a rise of anti-EU voting, that have lead to, for example, the United Kingdom's 2016 vote to exit the European Union. A number of studies explore correlates with, and potential origins of, support for Eurosceptic political parties \citep[for example][]{brookings_eu, becker_2017, colantone_stanig_2018, inglehart_2016}. While much of this research emphasizes either economic insecurity or a cultural backlash against multiculturalism/progressive values, a related strand of research explores the role of personal connections in shaping political preferences. Early evidence was provided by \cite{lazarsfeld_1944}, who  documents the influence of friends on U.S. voters. More recently, \cite{mclaren_2003} finds that intimate contact with members of minority groups can reduce individuals' willingness to expel immigrants and \cite{algan_2019} finds that the formation of friendships between students with different political opinions causes their views to converge. Motivated by this literature, we explore whether  differences in how socially connected a region is to other European countries is associated with higher trust in the European Union and a reduced propensity to vote for Eurosceptic political parties.

\begin{figure}[htb]
  \caption{International Connectedness within Europe}
  \label{fig:international_shares}
  \makebox[\textwidth][c]{\includegraphics[width=0.95\textwidth]{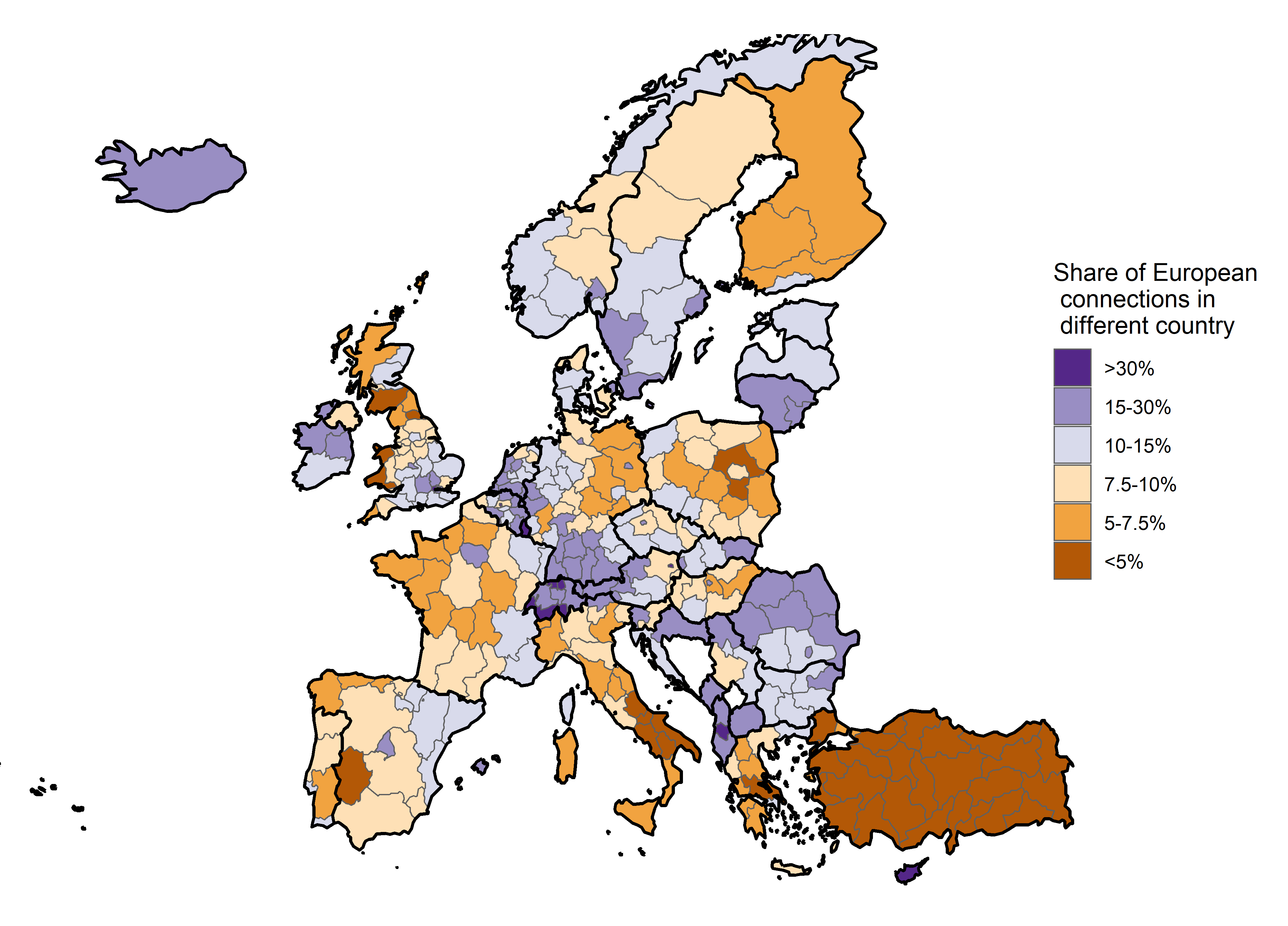}}
    \begin{footnotesize}
        \textit{Note:} Figure shows the share of European friendship links from users in each region to users in a different country. Darker orange regions have a lower share of European friends in different countries and darker purple regions have a higher share.
  \end{footnotesize}
\end{figure}

European regions vary considerably on the share of social connections to individuals living in foreign countries.\footnote{In our data, we focus on connections between individuals living within Europe. Therefore, (as we describe) our measure is the share of $European$ connections to individuals in different $European$ countries, rather the the share of $all$ connections in $any$ different country.} For the median NUTS2 region, 9.8\% of European connections are to individuals in different countries. However, in the least internationally connected decile of regions, less than 4.1\% of connections are to individuals living in different countries, compared to over 19.7\% in the most internationally connected decile. Figure \ref{fig:international_shares} maps this share for every region.

\paragraph*{Regional Data on Euroscepticism.} We analyze two measures of  Euroscepticism. Our first measure is a survey measure of trust in the EU. As part of the  ``Eurobarometer'' series of public opinion surveys undertaken for the European Commission, the \textit{Flash Eurobarometer 472: Public opinion in the EU regions} asks respondents about their trust in the EU. Specifically, the survey asks respondents ``I would like to ask you a question about how much trust you have in the European Union. Could you tell me if you tend to trust it or tend not to trust it?'' \citep{eurobarometer}. The survey was conducted between October and November 2018, and had 61,968 respondents (around 300 per region). Second, we use the vote share for anti-EU parties. Regional-level data on parliamentary and presidential elections in Europe from 2000 to 2017 are available from \cite{brookings_eu}.\footnote{Data and programs are available at: \url{https://www.brookings.edu/bpea-articles/the-european-trust-crisis-and-the-rise-of-populism/}.} As described in the paper, the determination of whether a particular party is ``Anti-EU'' is primarily based on the Chapel Hill Expert Survey, with some additional analysis of party platforms by the paper's authors. From this data, we use the average share of the electorate that voted for Anti-EU parties in elections between 2009 and 2017. In addition to these two measures of Euroscepticism, we use demographic and socioeconomic data from Eurostat, including the share of each region's population that was born in a different European country.

\paragraph*{Relationship between Euroscepticism and Connectedness.} To systematically explore the relationship between regional views on the European Union and international friendships, we use the equation:
\begin{equation}
\begin{split}
    \label{eq:reg_eu}
    Share\_EU\_View_{i} = \beta_0*Share\_Connections\_Foreign_{i} + X_{i} + \zeta_{c(i)} + \epsilon_{i}
\end{split}
\end{equation}
Here, $Share\_EU\_View_{i}$ is either the share of individuals responding that they trust the European Union in the Eurobarometer survey or the average share of the electorate that voted for Anti-EU parties between 2009 and 2017. Our unit of observation for this regression is the level at which we observe these two outcomes of interest. For both data sets, this includes a mixture of NUTS1 and NUTS2 regions. The first explanatory variable, $Share\_Connections\_Foreign_{i}$, is the share of the region's European connections that are to individuals in foreign countries. $X_i$ is a set of regional socioeconomic characteristics. These are average income, unemployment rate, and the shares of employment in manufacturing, construction, and professional sectors. They may also include the share of residents living in the region who are born in other European countries. The controls are indicators based on the deciles of each measure, to capture any non-linear relationships. Some specifications include country fixed effects, denoted here by $\zeta_{c(i)}$.

Table \ref{tab:eu_views} shows the results of this regression. Columns 1 and 5 show the relationships between the outcomes of interest and the share of the region's European connections that are to individuals in foreign countries: a 1 percentage point increase in the share of connections that are to other European countries is associated with a 0.50 percentage point increase in the share of residents who trust the EU and a 0.76 percentage point decrease in the share of votes for an Anti-EU political party. The second and fifth columns show that when adding the socioeconomic controls listed above, the magnitude of the relationship between the trust in the EU decreases and becomes insignificant, but the magnitude of the relationship with anti-EU voting slightly increases. When we also add the share of the population that is born in other European countries (columns 3 and 7), the magnitude of the relationship with trust increases and the magnitude of the relationship with voting decreases. There is a complicated relationship between demographic and socioeconomic factors, and views on the European Union; however, the directionally persistent relationship between Euroscepticism and international social connectedness at the regional level suggest foreign connections might play a role in shaping views on the EU, a result that would be consistent with the existing literature on personal connections impacting political preferences. Columns 4 and 8 show that, after adding country-level fixed effects, increases in a region's foreign connections continue to imply a decrease in the share of residents that trust the EU, while th effect on the share that vote for anti-EU parties becomes insignificant. This suggests that country-specific factors (e.g. the organizational resources of anti-EU political parties within the country) dominate possible effects of international connections in shaping Eurosceptic voting patterns, but not in shaping individuals' views.

\begin{table}
\begin{threeparttable}[h]
    \begin{singlespacing}
    \begin{footnotesize}
    \caption{International Connections and Views on the European Union}
    \label{tab:eu_views}
    \footnotesize
        {
        \def\sym#1{\ifmmode^{#1}\else\(^{#1}\)\fi}
        \begin{tabularx}{1.015\textwidth}{l*{8}{>{\centering\arraybackslash} X }}
            \toprule\toprule
                        &\multicolumn{4}{c}{Share Trust in EU (\%)}                                             &\multicolumn{4}{c}{Share Vote for Anti-EU Parties (\%)}                                \\
                        \cmidrule(l){2-5} \cmidrule(l){6-9}
                        &\multicolumn{1}{c}{(1)}         &\multicolumn{1}{c}{(2)}         &\multicolumn{1}{c}{(3)}         &\multicolumn{1}{c}{(4)}         &\multicolumn{1}{c}{(5)}         &\multicolumn{1}{c}{(6)}         &\multicolumn{1}{c}{(7)}         &\multicolumn{1}{c}{(8)}         \\
            \midrule
            Share of European friends&       0.499\sym{***}&       0.161         &       0.254\sym{*}  &       0.326\sym{**} &      -0.762\sym{***}&      -0.819\sym{***}&      -0.541\sym{**} &       0.454         \\
            outside of country (\%)&     (0.101)         &     (0.127)         &     (0.153)         &     (0.155)         &     (0.187)         &     (0.230)         &     (0.273)         &     (0.303)         \\
                        &                     &                     &                     &                     &                     &                     &                     &                     \\
            Socioeconomic Controls&                    &           Y         &           Y         &           Y         &                    &           Y         &           Y         &           Y         \\
            Share Foregin Born Controls&                    &                    &           Y         &           Y         &                    &                    &           Y         &           Y         \\
            Country Fixed Effects&                    &                    &                    &           Y         &                    &                    &                    &           Y         \\
            \midrule
            $R^2$       &       0.109         &       0.503         &       0.571         &       0.891         &       0.071         &       0.493         &       0.568         &       0.818         \\
            Number of Observations&         201         &         201         &         201         &         201         &         221         &         221         &         221         &         221         \\
	\bottomrule\bottomrule
        \end{tabularx}
        }
        \begin{tablenotes}[flushleft]
        \item[] \textit{Note:} Table shows results from Regression \ref{eq:reg_eu}. The dependent variable in columns 1-4 is the share of individuals responding that they trust the European Union in a survey conducted for the European Commission. The dependent variable in columns 5-8 is the average share of the electorate the voted for ``Anti-EU'' parties between 2009 and 2017. The unit of observation is the level at which we observe each dependent variable (either a NUTS2 or NUTS1 region). Columns 1 and 5 include only one explanatory variable: the share of the region's European connections that are to individuals in foreign countries. Columns 2 and 6 add a set of demographic and economic controls. Columns 3 and 7 add controls for the share of the region's population that was born in a different country. Columns 4 and 8 add country-level fixed effects. Significance levels: *(p$<$0.10), **(p$<$0.05), ***(p$<$0.01).
    \end{tablenotes}
    \end{footnotesize}
    \end{singlespacing}
\end{threeparttable}
\end{table}

\section{Additional Data Details}

\subsection{Facebook and Real World Connections.} 
Establishing a Facebook friendship link requires the consent of both individuals, and the total number of friends for a person is limited to 5,000. As a result, Facebook connections are primarily between real-world acquaintances. Indeed, one independent survey of American Facebook users revealed that only 39\% of users reported being Facebook friends with someone that they have never met in person \citep{pew2015}. In contrast, Facebook users reported that they were generally Facebook friends with individuals for which they have real-world connections: 93\% said they were Facebook friends with family other than parents or children, 91\% said they were connected with current friends, 87\% said they were connected to past friends (such as former classmates), and 58\% said they were connected to work colleagues \citep{pew2015}. As a result, networks formed on Facebook more closely resemble real-world social networks than those on other online platforms, such as Twitter, where uni-directional links to non-acquaintances, such as celebrities, are common. In prior work Facebook friendships have been shown to be useful to describe real world networks. For example, social connectedness as measured through Facebook friendship links is strongly related to patterns of COVID-19 spread \citep{kuchler2020_covid}, international trade \citep{bailey2020_international}, and investment decisions \citep{kuchler2020_institution}. See \cite{bailey2019house, Bailey2018a, Bailey2018b, Bailey2019a, Bailey2019b} for additional discussion of the evidence that friendships observed on Facebook serve as a good proxy for real-world social connections.

\subsection{Subset of Friendships.} 
As one way of better understanding the connections underlying our measure of Social Connectedness in the primary paper, we compare it to similar measures constructed using restricted sets of friendships. Table \ref{tab:scis_correlation} presents the cross-correlation of measures limited to connections made during certain periods of time (e.g., recent friendships vs old friendships) and to friendships between individuals with certain shared characteristics (e.g., ages less than 5 years apart). Each of these measures is highly correlated with the others. This provides more evidence that Facebook connections resemble full real-world networks and not, for example, primarily recently formed online-only connections.

\begin{table}[htbp]
    \begin{singlespacing}
    \caption{Correlation of Social Connectedness Constructed from Select Friendship Pairs}
    \footnotesize
    \label{tab:scis_correlation}
   \begin{tabularx}{\textwidth}{l*{8}{>{\centering\arraybackslash} X }}
  \toprule
  \toprule
 & (1) & (2) & (3) & (4) & (5) & (6) & (7) & (8) \\ 
  \midrule
  (1) All & 1.000 &  &  &  &  &  &  &  \\ 
  [0.5em]
  (2) Added $<$1 Year Ago & 0.946 & 1.000 &  &  &  &  &  &  \\ 
  [0.5em]
  (3) Added 1-5 Years Ago & 0.991 & 0.967 & 1.000 &  &  &  &  &  \\ 
  [0.5em]
  (4) Added $<$5 Years Ago & 0.985 & 0.982 & 0.998 & 1.000 &  &  &  &  \\ 
  [0.5em]
  (5) Added $>$5 Years Ago & 0.970 & 0.848 & 0.932 & 0.915 & 1.000 &  &  &  \\ 
  [0.5em]
  (6) Both Female & 0.982 & 0.910 & 0.960 & 0.953 & 0.975 & 1.000 &  &  \\ 
  [0.5em]
  (7) Both Male & 0.986 & 0.940 & 0.982 & 0.977 & 0.949 & 0.955 & 1.000 &  \\ 
  [0.5em]
  (8) Ages Within 5 Years & 0.997 & 0.942 & 0.982 & 0.977 & 0.974 & 0.983 & 0.979 & 1.000 \\ 
  \bottomrule
  \bottomrule
    \end{tabularx}
    \begin{footnotesize}
    \begin{tablenotes}[flushleft]
    \item[] \textit{Note:} Table presents correlations between social connectedness and similar measures constructed from restricted sets of connections.  Rows and Columns 2-5 limit to connections made less than a year ago, between 1 and 5 years ago, less than five years ago, and more than 5 years ago. Rows and Columns 5-8 limit to connections between females, males, and individuals with ages 5 or fewer years apart.  \\ \\
    \end{tablenotes}
    \end{footnotesize}
    \end{singlespacing}
\end{table}

\subsection{Regional Demographic and Socioeconomic Data.}
Information on demographic and socioeconomic characteristics of each region, such as educational attainment, median age, average income, and unemployment rate, is available from Eurostat. We calculate a measure of region-to-region industrial composition similarity using employment data from Eurostat's Structural Business Statistics series. For regional data on language and religion, we use the European Social Survey. In particular, the survey asks respondents which language they speak most often at home and --- if the respondent considers him- or herself religious --- their religious affiliation \citep{ESS}.\footnote{Most of our data come from Wave 8 of the survey, which was conducted in 2016. However, when a country was not included in Wave 8, we use the most recent wave of the survey for which the country was included. These countries are Denmark (Wave 7, 2014), Albania, Bulgaria, Cyprus and Slovakia (Wave 6, 2012), Croatia and Greece (Wave 5, 2010), and Latvia and Romania (Wave 4, 2008). In addition, Malta was not surveyed but is comprised of a single NUTS2 region. According to other survey data, 97\% of the population considers Maltese their ``mother tongue'' \citep{eurobarometer_language} and 95\% identifies as Roman Catholic \citep{malta_today}. We include these as the most common language and religion.} Our analysis focuses on the regional pairs within the smaller set of countries for which we have a full set of control data.\footnote{We do not include Albania, Iceland, Lichtenstein, Luxembourg, Montenegro, North Macedonia, Serbia, Switzerland, and Turkey; most of these countries are excluded because they are not part of the European Union, and therefore do not participate in many of the data collection efforts we use to construct our data.}

\subsection{Passenger Train Travel Data} \label{appendix:pass_train}
Our analyses in Section \ref{sec:travel} use information on region-to-region passenger train travel made available by Eurostat in the series ``tran\_r\_rapa.'' The data are based on individual reports from Member States of the European Union or European Free Trade Association, as well as European Union Candidate Countries. For each observation, the reporting country is identified. The data are reported for 2005, 2010, and 2015, in accordance with Regulation 91/2003 of the European Commission and subsequent regulatory updates. We take a number of steps to clean the data to prepare it for our analyses. This process was informed by both the ``Reference Manual on Rail transport statistics'' \citep{eurostat_train} and correspondence with the Eurostat data providers.

We first restrict our data to observations at the NUTS2 level, removing any country-level observations (we do, however, keep country-level data for countries which have only a single NUTS2 region). We also exclude all pairs that include a region with the unknown indicator ``XX'' or the extraregio territory indicator ``ZZ.'' These pairs make up around 1.3\% of passenger trips in the data. From here, we are faced with four challenges: 1) As confirmed by the authors' correspondence with Eurostat, when the data appear as ``non-available'' in a particular row this could mean either that there was no traffic or that the relevant country did not provide the data.\footnote{In some instances, countries report the data to Eurostat, but flag them as confidential so that they are not included in the public release. We always treat these data as missing in our final analysis.} 2) There are a number of hypothetical regional pairs missing, even between countries that did report data elsewhere. 3) For some international regional pairs, there are data reported from both countries on the same train flows, and often the number of passengers does not match. 4) NUTS2 classifications changed in 2006, 2010, and 2013. Each year of data is reported using the NUTS2 classification that was relevant at that particular point in time.

\begin{table}
\begin{threeparttable}[h]
    \begin{singlespacing}
    \begin{footnotesize}
    \caption{Passenger train travel data availability }
    \label{tab:appendix_train}
    \footnotesize
    \begin{tabularx}{\textwidth}{l*{7}{>{\centering\arraybackslash} X }} \toprule\toprule
        reporter & domestic2015 & international2015 & domestic2010 & international2010 & domestic2005 & international2005 \\ \midrule
        AL & 0 & 0 & 0 & 0 & 0 & 0 \\
        AT & 0 & 0 & 0 & 0 & 0 & 0 \\
        BE & 0 & 0 & 1 & 1 & 1 & 1 \\
        BG & 1 & 1 & 1 & 1 & 0 & 0 \\
        CH & 1 & 1 & 1 & 1 & 0 & 0 \\
        CY & 0 & 0 & 0 & 0 & 0 & 0 \\
        CZ & 1 & 1 & 1 & 1 & 1 & 1 \\
        DE & 1 & 0 & 1 & 0 & 1 & 1 \\
        DK & 0 & 0 & 0 & 0 & 1 & 1 \\
        EE & 1 & 0 & 1 & 0 & 1 & 0 \\
        EL & 0 & 0 & 1 & 0 & 1 & 0 \\
        ES & 1 & 0 & 1 & 0 & 1 & 0 \\
        FI & 1 & 0 & 1 & 0 & 1 & 0 \\
        FR & 0 & 0 & 0 & 0 & 0 & 0 \\
        HR & 1 & 1 & 1 & 1 & 0 & 0 \\
        HU & 0 & 0 & 1 & 1 & 1 & 1 \\
        IE & 1 & 1 & 1 & 1 & 1 & 1 \\
        IS & 0 & 0 & 0 & 0 & 0 & 0 \\
        IT & 1 & 0 & 0 & 0 & 1 & 0 \\
        LI & 0 & 0 & 0 & 0 & 0 & 0 \\
        LT & 1 & 1 & 1 & 1 & 1 & 1 \\
        LU & 1 & 1 & 1 & 1 & 1 & 1 \\
        LV & 1 & 1 & 1 & 1 & 1 & 1 \\
        ME & 0 & 0 & 0 & 0 & 0 & 0 \\
        MK & 0 & 0 & 0 & 0 & 0 & 0 \\
        MT & 0 & 0 & 0 & 0 & 0 & 0 \\
        NL & 0 & 0 & 0 & 0 & 1 & 1 \\
        NO & 1 & 1 & 1 & 0 & 1 & 0 \\
        PL & 1 & 1 & 1 & 1 & 1 & 1 \\
        PT & 1 & 0 & 1 & 0 & 1 & 0 \\
        RO & 0 & 0 & 0 & 0 & 0 & 0 \\
        RS & 0 & 0 & 0 & 0 & 0 & 0 \\
        SE & 0 & 0 & 0 & 0 & 0 & 0 \\
        SI & 1 & 1 & 1 & 1 & 1 & 1 \\
        SK & 1 & 1 & 1 & 1 & 1 & 1 \\
        TR & 1 & 0 & 1 & 0 & 0 & 0 \\
        UK & 0 & 0 & 0 & 0 & 0 & 1 \\ \bottomrule\bottomrule
    \end{tabularx}
    \begin{tablenotes}[flushleft]
    \item[] \textit{Note:} Table shows the regional passenger train travel data availability by reporting country, year, and whether the travel is domestic or international. 0s indicate the data were not available and 1s indicate the data were available. Reporting country is given by the two-letter prefix of each country's NUTS codes. The table only shows whether any data from a particular \textit{reporter} were available, not whether any regions from this country are included in the final analysis. For example, although Austria did not report international data in 2015, pairs that include an Austrian region and a region in a country that did report international data in 2015 are included.
    \end{tablenotes}
    \end{footnotesize}
    \end{singlespacing}
\end{threeparttable}
\end{table}

With respect to challenges 1 and 2, we found that each country reports data to Eurostat in two intermediate data sets: one for domestic passenger travel and another for international passenger travel. To identify countries that submitted a particular set of data in a particular year, we group the data by the reporting country, year, and whether the region pair is international or domestic. We then generate a list of countries that had at least one non-missing entry in each year/domestic-international group. These lists are provided in Table \ref{tab:appendix_train}.  When ``non-availble'' values are reported by a country that \textit{did not} report data elsewhere in the year/domestic-international group, we treat this as missing and exclude it. When ``non-available'' values are reported by a country that \textit{did} report data elsewhere in the group, we treat this value as a zero (no traffic). Additionally, for countries that reported data in a particular group, we fill any missing regional pairs (i.e. pairs that are not in the data) in the group with zeros. Together, these assumptions handle challenges 1 and 2.

For each international regional pair, there still remains two possible reports: one from each of the regions' home countries in the pair. In instances when only one country reports the data, we take the non-missing value from the reporting country. However, there are a number of instances when each country reports data for the same international regional pair (challenge 3). In these instances, we take the average of the two reports. Finally, to update the data to the 2016 NUTS2 regions (challenge 4) we build a crosswalk using the history of NUTS information provided by Eurostat.\footnote{Available at: \url{https://ec.europa.eu/eurostat/web/nuts/history}} In instances when an older NUTS2 region split into multiple new regions, we set the number of passengers in each row that includes a new region equal to the corresponding old row's number of passengers, multiplied by the new region's population share of the old region population (i.e. we assume that passenger train travel in each of these regions is proportional to population).

\section{Historical Europe Maps} \label{appendix:history_maps}

Analyses in our primary paper use information on the country that each modern NUTS2 region was a part of in the years 1900, 1930, 1960, and 1990. The maps in this appendix show these country classifications for each year. The data largely come from files provided by the Max Planck Institute for Demographic Research Population History GIS Collection \citep{hist_maps}. In cases when a modern region spans two historical countries, we classify the region as part of the country for which it had a greater land area overlap.

\begin{figure}[hp]
  \caption{2019 countries of Europe}
  \label{fig:countries_2019}
  \makebox[\textwidth][c]{\includegraphics[width=0.95\textwidth]{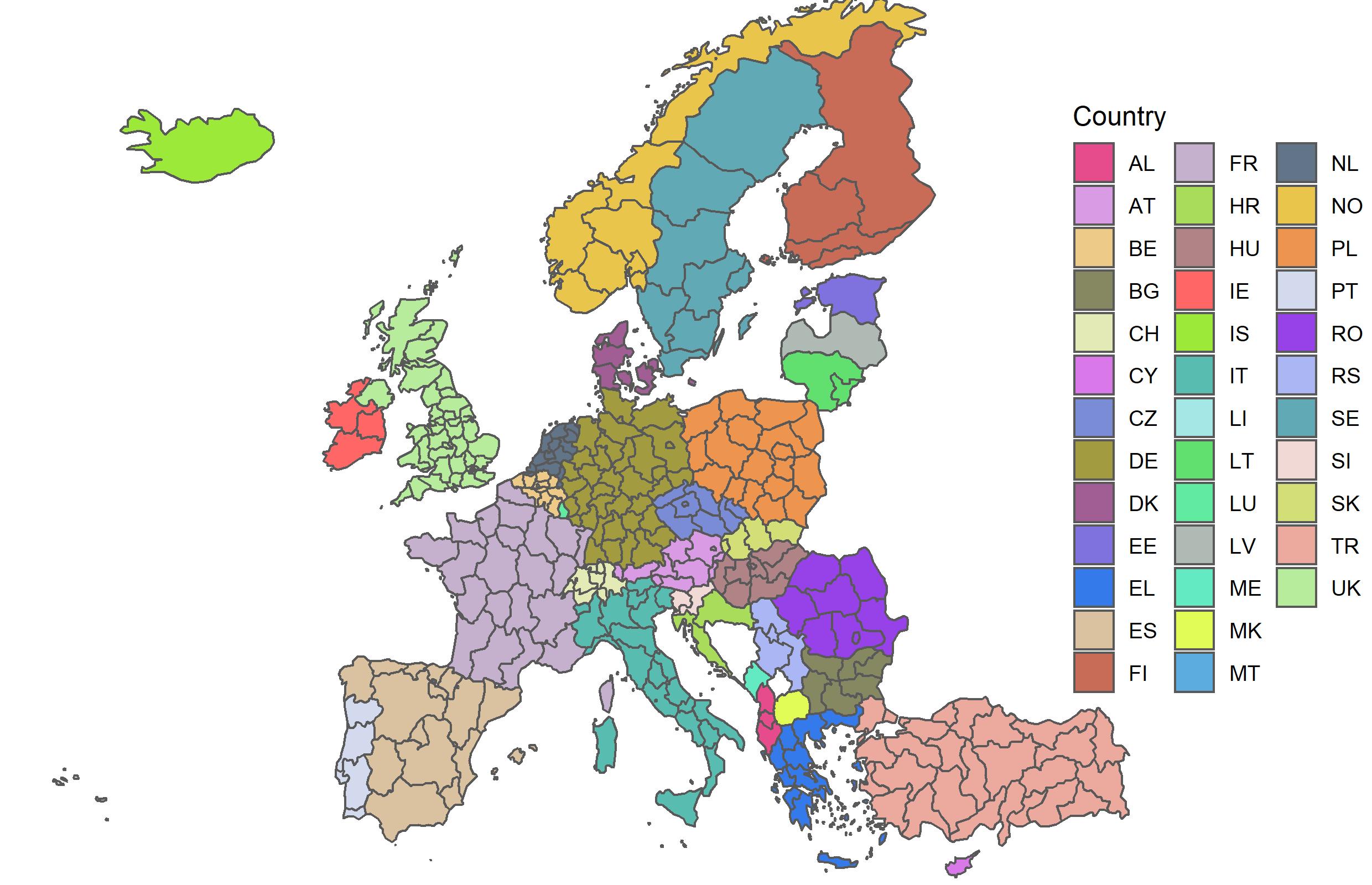}}
    \begin{footnotesize}
        \textit{Note:} Figure shows the 2019 country of each present-day NUTS2 region.
  \end{footnotesize}
\end{figure}

\begin{figure}[hp]
  \caption{1990 and 1960 countries of Europe}
  \label{fig:countries_1990_1960}
  \makebox[\textwidth][c]{\includegraphics[width=0.95\textwidth]{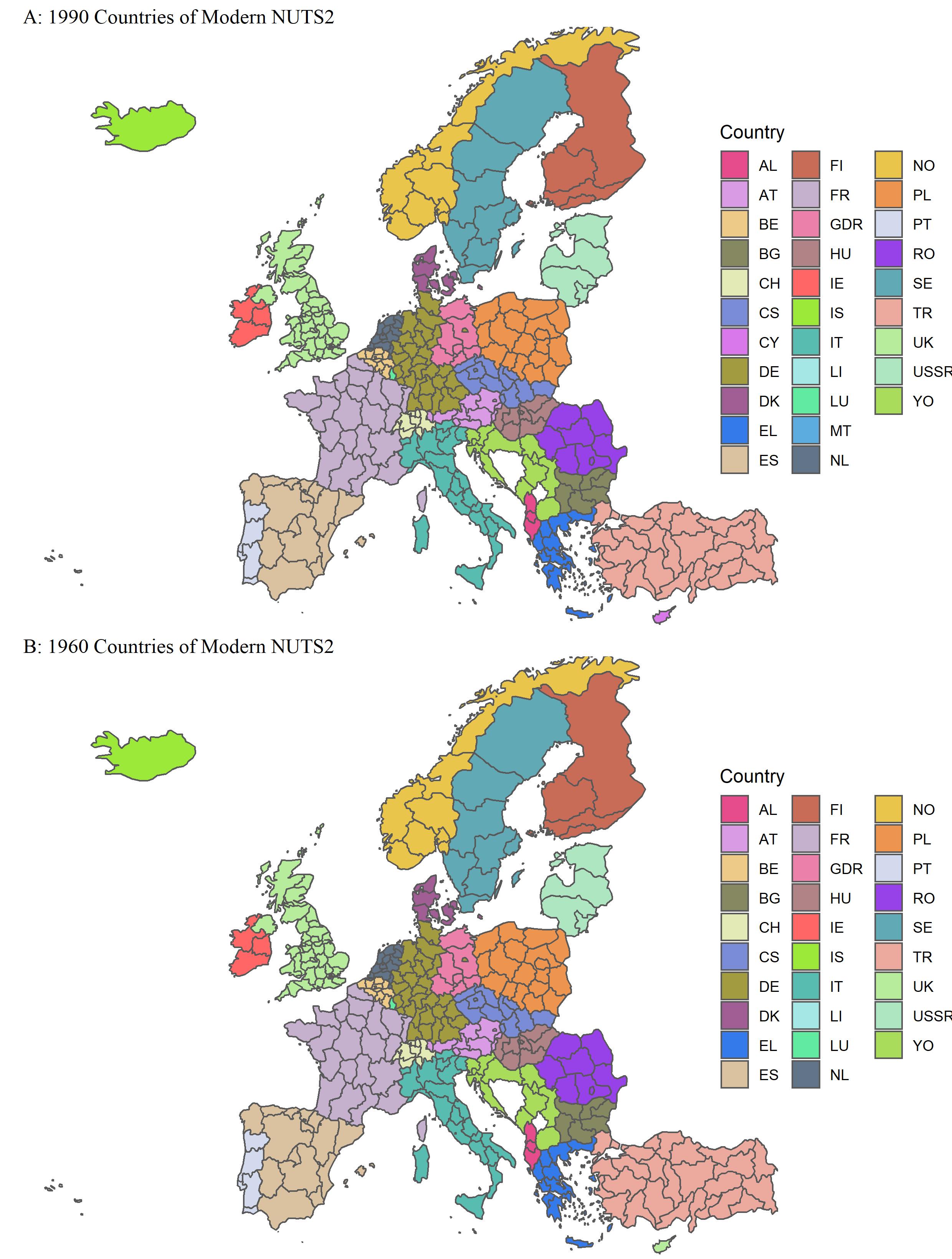}}
    \begin{footnotesize}
        \textit{Note:} Figure shows the 1990 country (Panel A) and 1960 country (Panel B) of each present-day NUTS2 region.
  \end{footnotesize}
\end{figure}

\begin{figure}[hp]
  \caption{1930 and 1900 countries of Europe}
  \label{fig:countries_1900_1930}
  \makebox[\textwidth][c]{\includegraphics[width=0.95\textwidth]{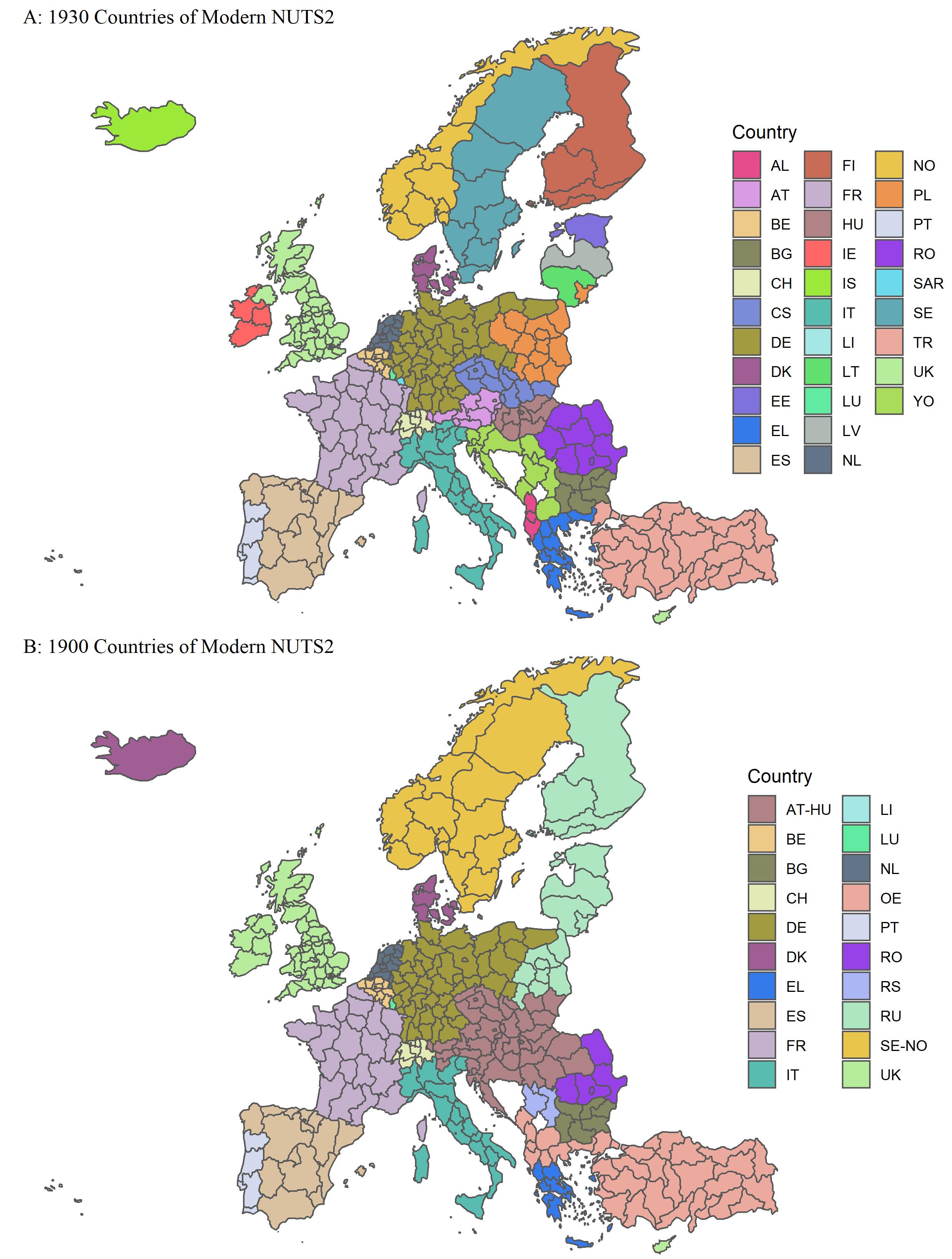}}
    \begin{footnotesize}
        \textit{Note:} Figure shows the 1930 country (Panel A) and 1900 country (Panel B) of each present-day NUTS2 region.
  \end{footnotesize}
\end{figure}

\clearpage
\bibliography{bib}

\begin{thebibliography}{33}
\providecommand{\natexlab}[1]{#1}
\providecommand{\url}[1]{\texttt{#1}}
\expandafter\ifx\csname urlstyle\endcsname\relax
  \providecommand{\doi}[1]{doi: #1}\else
  \providecommand{\doi}{doi: \begingroup \urlstyle{rm}\Url}\fi

\bibitem[Algan et~al.(2017)Algan, Guriev, Papaioannou, and
  Passari]{brookings_eu}
Yan Algan, Sergei Guriev, Elias Papaioannou, and Evgenia Passari.
\newblock The european trust crisis and the rise of populism.
\newblock \emph{Brookings Papers on Economic Activity}, 2017.

\bibitem[Algan et~al.(2019)Algan, Dalvit, Do, Chapelain, and Zenou]{algan_2019}
Yann Algan, Nicol{\`o} Dalvit, Quoc-Anh Do, Alexis~Le Chapelain, and Yves
  Zenou.
\newblock Friendship networks and political opinions: A natural experiment
  among future french politicians.
\newblock Working Paper 1294, Research Institute of Industrial Economics, 2019.

\bibitem[Axhausen(2008)]{axhausen2008networks}
K~W Axhausen.
\newblock Social networks, mobility biographies, and travel: Survey challenges.
\newblock \emph{Environment and Planning B: Planning and Design}, 35\penalty0
  (6):\penalty0 981--996, 2008.

\bibitem[Aydin(2016)]{turkishMigration}
Yasar Aydin.
\newblock The germany-turkey migration corridor: Refitting policies for a
  transnational age.
\newblock Report, Migration Policy Institute, 2016.

\bibitem[Bailey et~al.(2018{\natexlab{a}})Bailey, Cao, Kuchler, and
  Stroebel]{Bailey2018a}
Michael Bailey, Rachel Cao, Theresa Kuchler, and Johannes Stroebel.
\newblock The economic effects of social networks: Evidence from the housing
  market.
\newblock \emph{Journal of Political Economy}, 126\penalty0 (6):\penalty0
  2224--2276, 2018{\natexlab{a}}.

\bibitem[Bailey et~al.(2018{\natexlab{b}})Bailey, Cao, Kuchler, Stroebel, and
  Wong]{Bailey2018b}
Michael Bailey, Rachel Cao, Theresa Kuchler, Johannes Stroebel, and Arlene
  Wong.
\newblock Social connectedness: Measurements, determinants, and effects.
\newblock \emph{Journal of Economic Perspectives}, 32\penalty0 (3):\penalty0
  259--80, 2018{\natexlab{b}}.

\bibitem[Bailey et~al.(2018{\natexlab{c}})Bailey, Dávila, Kuchler, and
  Stroebel]{bailey2019house}
Michael Bailey, Eduardo Dávila, Theresa Kuchler, and Johannes Stroebel.
\newblock {House Price Beliefs And Mortgage Leverage Choice}.
\newblock \emph{The Review of Economic Studies}, 86\penalty0 (6):\penalty0
  2403--2452, 11 2018{\natexlab{c}}.
\newblock ISSN 0034-6527.
\newblock \doi{10.1093/restud/rdy068}.
\newblock URL \url{https://doi.org/10.1093/restud/rdy068}.

\bibitem[Bailey et~al.(2019{\natexlab{a}})Bailey, Farrell, Kuchler, and
  Stroebel]{Bailey2019a}
Michael Bailey, Patrick Farrell, Theresa Kuchler, and Johannes Stroebel.
\newblock Social connectedness in urban areas.
\newblock Working Paper 26029, National Bureau of Economic Research,
  2019{\natexlab{a}}.

\bibitem[Bailey et~al.(2019{\natexlab{b}})Bailey, Johnston, Kuchler, Stroebel,
  and Wong]{Bailey2019b}
Michael Bailey, Drew~M Johnston, Theresa Kuchler, Johannes Stroebel, and Arlene
  Wong.
\newblock Peer effects in product adoption.
\newblock Working Paper 25843, National Bureau of Economic Research,
  2019{\natexlab{b}}.

\bibitem[Bailey et~al.(2020)Bailey, Gupta, Hillenbrand, Kuchler, Richmond, and
  Stroebel]{bailey2020_international}
Michael Bailey, Abhinav Gupta, Sebastian Hillenbrand, Theresa Kuchler, Robert
  Richmond, and Johannes Stroebel.
\newblock International trade and social connectedness.
\newblock Working Paper 26960, National Bureau of Economic Research, 2020.

\bibitem[Becker et~al.(2017)Becker, Fetzer, and Novy]{becker_2017}
Sascha~O Becker, Thiemo Fetzer, and Dennis Novy.
\newblock {Who voted for Brexit? A comprehensive district-level analysis}.
\newblock \emph{Economic Policy}, 32\penalty0 (92):\penalty0 601--650, 2017.

\bibitem[Carrasco and Miller(2009)]{carrasco2009social}
Juan-Antonio Carrasco and Eric~J. Miller.
\newblock The social dimension in action: A multilevel, personal networks model
  of social activity frequency between individuals.
\newblock \emph{Transportation Research Part A: Policy and Practice},
  43\penalty0 (1):\penalty0 90--104, 2009.

\bibitem[Colantone and Stanig(2018)]{colantone_stanig_2018}
Italo Colantone and Piero Stanig.
\newblock Global competition and brexit.
\newblock \emph{American Political Science Review}, 112\penalty0 (2):\penalty0
  201–218, 2018.

\bibitem[Correia et~al.(2019)Correia, Guimaraes, and Zylkin]{Correia}
Sergio Correia, Paulo Guimaraes, and Tom Zylkin.
\newblock Ppmlhdfe: Fast poisson estimation with high-dimenstional fixed
  effects, 2019.

\bibitem[Duggan et~al.(2015)Duggan, Ellison, Lampe, Lenhart, and
  Madden]{pew2015}
Maeve Duggan, Nicole~B Ellison, Cliff Lampe, Amanda Lenhart, and Mary Madden.
\newblock Social media update 2014.
\newblock Report, Pew Research Center, 2015.

\bibitem[European Comission({\natexlab{a}})]{eurobarometer_language}
European Comission.
\newblock Special eurobarometer 386: Europeans and their languages.
\newblock Report, 2012{\natexlab{a}}.

\bibitem[European Comission({\natexlab{b}})]{eurobarometer}
European Comission.
\newblock Flash eurobarometer 472: Public opinion in the eu regions.
\newblock Report, 2018{\natexlab{b}}.

\bibitem[European Commission()]{erasmus}
European Commission.
\newblock What is erasmus+?
\newblock URL \url{https://ec.europa.eu/programmes/erasmus-plus/about_en}.

\bibitem[European Social Survey()]{ESS}
European Social Survey.
\newblock European Social Survey Cumulative File, ESS 1-8 (2018). Data file
  edition 1.0. NSD - Norwegian Centre for Research Data, Norway - Data Archive
  and distributor of ESS data for ESS ERIC. doi:10.21338/NSD-ESS-CUMULATIVE.

\bibitem[European Union()]{eu_goals}
European Union.
\newblock The eu in brief: Goals and values of the eu.
\newblock URL
  \url{https://europa.eu/european-union/about-eu/eu-in-brief_en#goals-and-values-of-the-eu}.

\bibitem[Eurostat()]{eurostat_train}
Eurostat.
\newblock Reference manual on rail transport statistics version 10.1.
\newblock Technical report, 2019.

\bibitem[Harrison(2018)]{romaniaLondon}
George Harrison.
\newblock Inside little bucharest: How romanian immigration changed this small
  london suburb beyond recognition.
\newblock \emph{The Sun}, 2018.

\bibitem[Inglehart and Norris(2016)]{inglehart_2016}
Ronald Inglehart and Pippa Norris.
\newblock Trump, brexit, and the rise of populism: Economic have-nots and
  cultural backlash.
\newblock Working Paper RWP16-026, Harvard Kennedy School, August 2016.

\bibitem[Kim et~al.(2018)Kim, Rasouli, and Timmermans]{kim2018networks}
Jinhee Kim, Soora Rasouli, and Harry J.~P. Timmermans.
\newblock Social networks, social influence and activity-travel behaviour: a
  review of models and empirical evidence.
\newblock \emph{Transport Reviews}, 38\penalty0 (4):\penalty0 499--523, 2018.

\bibitem[Kuchler et~al.(2020{\natexlab{a}})Kuchler, Peng, Stroebel, Li, and
  Zhou]{kuchler2020_institution}
Theresa Kuchler, Lin Peng, Johannes Stroebel, Yan Li, and Dexin Zhou.
\newblock Social proximity to capital: Implications for investors and firms.
\newblock Technical report, 2020{\natexlab{a}}.
\newblock Working paper.

\bibitem[Kuchler et~al.(2020{\natexlab{b}})Kuchler, Russel, and
  Stroebel]{kuchler2020_covid}
Theresa Kuchler, Dominic Russel, and Johannes Stroebel.
\newblock The geographic spread of covid-19 correlates with structure of social
  networks as measured by facebook.
\newblock Working Paper 26990, National Bureau of Economic Research,
  2020{\natexlab{b}}.

\bibitem[Lazarsfeld et~al.(1944)Lazarsfeld, Berelson, and
  Gaudet]{lazarsfeld_1944}
Paul Lazarsfeld, Bernard Berelson, and Hazel Gaudet.
\newblock \emph{The People's Choice: How the Voter Makes up His Mind in a
  Presidential Campaign}.
\newblock Columbia University Press, 1944.

\bibitem[McLaren(2003)]{mclaren_2003}
Lauren~M. McLaren.
\newblock Anti-immigrant prejudice in europe: Contact, threat perception, and
  preferences for the exclusion of migrants.
\newblock \emph{Social Forces}, 81\penalty0 (3):\penalty0 909--936, 2003.

\bibitem[MPIDR and CGG()]{hist_maps}
MPIDR and CGG.
\newblock Max Planck Institute for Demographic Research and Chair for Geodesy
  and Geoinformatics, University of Rostock. MPIDR Population History GIS
  Collection – Europe (partly based on ©EuroGeographics for the
  administrative boundaries), 2013.

\bibitem[Páez and Scott(2007)]{paez2007social}
Antonio Páez and Darren~M Scott.
\newblock Social influence on travel behavior: A simulation example of the
  decision to telecommute.
\newblock \emph{Environment and Planning A: Economy and Space}, 39\penalty0
  (3):\penalty0 647--665, 2007.

\bibitem[Sansone(2018)]{malta_today}
Kurt Sansone.
\newblock Maltatoday survey: Maltese identity still very much rooted in
  catholicisim.
\newblock Report, 2018.

\bibitem[Schmidt(2017)]{immigration_croatia2017}
Alexandra Schmidt.
\newblock Why are croatians moving to ireland?
\newblock \emph{Croatia Week}, 2017.

\bibitem[Thomas(2019)]{immigration_croatia2019}
Mark Thomas.
\newblock Number of croatians moving to ireland increase tenfold.
\newblock \emph{The Dubrovnik Times}, 2019.

\end{thebibliography}
\bibliographystyle{plainnat}

\end{document}